\documentclass{aa}
\usepackage{graphicx}
\usepackage{rotating}
\begin{document}

   \title{{\em XMM-Newton} observation of the Lockman Hole
      \thanks{Based on observations obtained with {\em XMM-Newton}, an ESA 
        science 
           mission with instruments and contributions directly funded by 
           ESA Member States and the USA (NASA)}}

   \subtitle{II. Spectral analysis}

   \author{V. Mainieri\inst{1,2}
           \and J. Bergeron\inst{3}
           \and G. Hasinger\inst{4,5}
           \and I. Lehmann\inst{4}
           \and P. Rosati\inst{2}
           \and M. Schmidt\inst{6}
           \and G. Szokoly\inst{4,5}
           \and R. Della Ceca\inst{7}          
          }

   \offprints{V. Mainieri,\\ \email{vmainier@eso.org}}

   \institute{Dip. di Fisica, Universit\`a degli Studi Roma Tre, 
                Via della Vasca Navale 84, I-00146 Roma, Italy
        \and 
                European Southern Observatory,
                Karl-Schwarzschild-Strasse 2, D-85748 Garching, Germany
        \and
            Institut d'Astrophysique de Paris, 98bis Boulevard, F-75014 Paris, France
        \and
            Max-Planck-Institut f\"ur extraterrestrische Physik, Giessenbachstrasse PF 1312, 
                85748 Garching bei Muenchen, Germany
        \and
            Astrophysikalisches Institut, An der Sternwarte 16, Potsdam
                14482 Germany
        \and
            California Institute of Technology, Pasadena, CA $91125$, USA       
        \and 
            Osservatorio Astronomico di Brera, via Brera $28$, I-$20121$ Milano, Italy  
             }

   \date{Received 8 May 2002; accepted 2 July 2002 }

   \abstract{ We present the results of the X-ray spectral analysis of
   the first deep X-ray survey with the XMM-Newton observatory during
   Performance Verification.  The X-ray data of the Lockman Hole field
   and the derived cumulative source counts were reported by Hasinger
   et al. (2001).  We restrict the analysis to the sample of 98
   sources with more than 70 net counts (flux limit in the [0.5-7]
   keV band of $1.6 \times 10^{-15}$ erg cm$^{-2}$ s$^{-1}$) of which
   61 have redshift identification. We find no correlation between
   the spectral index $\Gamma$ and the intrinsic absorption column
   density N$_{\rm H}$ and, for both the Type-1 and Type-2 AGN
   populations, we obtain $\langle\Gamma\rangle \simeq 2$.  The
   progressive hardening of the mean X-ray source spectrum with decreasing
   flux is essentially due to an increase in intrinsic absorption.
   The marked separation between the two AGN populations in several
   diagnostics diagrams, involving X-ray colour, X-ray flux,
   optical/near IR colour and optical brightness, is also a
   consequence of different absorption column densities and enables the
   classification of optically faint obscured AGN. The Type-2
   and obscured AGN have weaker soft X-ray and optical fluxes and
   redder R$-$K$^\prime$ colours. They follow the evolutionary tracks
   of their host galaxies in a color-redshift diagram.  About 27$\%$
   of the subsample with R$-$K$^\prime$ colour are EROs (R$-$K$^\prime
   \geq 5$) and most of these 18 X-ray selected EROs contain an
   obscured AGN as revealed by their high X-ray-to-optical/near IR
   flux ratios.  There are six sources in our sample with ${\rm
   L_X}$[0.5-10]$>10^{44}$ erg s$^{-1}$ and ${\rm log(N_H)}>10^{22}$
   cm$^{-2}$: which are likely Type-2 QSOs and we thus derive 
   a density of $\sim 69$ objects of this class per square degree.
}


   \maketitle
%

\section{Introduction}

The deep ROSAT survey of the Lockman Hole showed that about 80$\%$ of
the soft (0.5-2 keV) X-ray background (XRB) is resolved into discrete
sources (Hasinger et al. \cite{gunther98}). These findings have
recently been confirmed and strengthened using the two deep Chandra
surveys of 1 Msec each (Brandt et al. \cite{brandt01}; Rosati et al.
\cite{piero02}). An important population of X-ray sources with hard
spectra, most probably obscured active galactic nuclei (AGN), is
present in the Chandra (Barger et al.
\cite{barger01}; Hornschemeier et al. \cite{hornschemeier01}; Rosati et al.  
\cite{piero02}) and XMM-Newton (Hasinger et al. \cite{paper1}, hereafter
Paper I) deep surveys; a few objects of this class had already been
detected in ROSAT deep and shallower surveys (Lehmann et
al. \cite{ingo01a}; Mittaz et al. \cite{mittaz99}). In the hard band
(2-10 keV), the X-ray source density derived from the number counts
in the two Chandra deep surveys is about 4000 deg$^{-2}$ (Brandt et
al. \cite{brandt01}; Rosati et al. \cite{piero02}) resolving $\sim
85-90 \%$ of the 2-10 keV XRB. This population of X-ray sources show a 
progressive hardening of the average X-ray spectrum towards fainter
fluxes (Tozzi et al. \cite{paolo01}; Mittaz et
al. \cite{mittaz99}).

The XMM-Newton deep survey ($\simeq$ 100 ksec of good quality data) of
the Lockman Hole was obtained during Performance Verification. The
X-ray data reduction and analysis (restricted to sources within a 10
arcmin radius) was reported in Paper I where it was demonstrated that
the different populations of X-ray sources are well separated in X-ray
spectral diagnostics based on hardness ratios. The extensive optical
follow-up programs of this field (Lehmann et al. \cite{ingo01a}, and
references therein) provide an understanding of the physical nature of
the X-ray sources.
The point sources detected in the soft band by ROSAT are 
predominantly unobscured (in both optical and X-ray bands) AGN 
spanning a wide redshift range.  In the XMM-Newton sample, there is a
significant fraction of sources with hard spectra. This new
population is most probably dominated by intrinsically absorbed AGN.
This assumption can be tested using the available optical
spectra and, more efficently, by X-ray spectral study.

To this aim, we have performed an X-ray spectral analysis of the
sources in the Lockman Hole to understand their physical nature
combining the X-ray data with the optical/near IR information. We also
use the subsample with redshift identification to check the validity
of our conclusions concerning the specific properties of the obscured
AGN population. Preliminary results of this work were reported by
Mainieri et al. (\cite{mainieri02}).\\ In the following we will refer
to Type-1 (broad and narrow emission lines) and Type-2 AGN (high
ionization narrow emission lines) using the optical spectroscopic
classification.

The observations are presented in Sect. 2. The results of the spectral
analysis are described in Sect. 3, in particular the range of the
X-ray spectral index, the observed ${\rm N_H}$ distribution and
colour-colour diagnostics diagrams. The optical/near IR properties are
discussed in Sect. 4 together with a comparison with QSO and galaxy
evolutionary tracks. The search for relations between X-ray and
optical/near IR fluxes is presented in Sect. 5. The effect of the
absorbing column density on the X-ray luminosity and the Type-2 QSO
candidates are discussed in Sect. 6.  Representative spectra of the
different classes of X-ray sources are given in Sect. 7. Finally, our
conclusions are outlined in Sect. 8.

\section{X-ray observations}

\subsection{Sample definition}

The X-ray results reported in this paper are obtained from the XMM
observation of the Lockman Hole field, centered on the sky position RA
10:52:43, DEC +57:28:48 (J2000). This is a region of extremely
low Galactic Hydrogen column density, 
${\rm N_H}=5.7 \times 10^{19}$ ${\rm cm}^{-2}$ (Lockman et al. 
\cite{lockman86}). The observation was performed in five separate
revolutions ($70$, $71$, $73$, $74$ and $81$) during the period April
$27$-May $19$, $2000$ for a total exposure time of $190$ ksec. Due to
periods of high background and flares, the good time intervals added up
to approximately $100$ ksec. 
 
The dataset, the cleaning procedure used, the source detection and the 
astrometric corrections are described in details in Paper I.
  
In this work, we use a sample of 192 sources with a likelihood value
$>10$ (corresponding to $\sim\!4\sigma$; see Paper I), and extend the
analysis to the whole Lockman Hole field of view (in Paper I only
sources with off-axis angle $<10\arcmin$ were considered). The flux
limits of this sample in the [0.5-2], [2-10] and [5-10] keV bands are
0.31, 1.4 and $2.4 \times 10^{-15}$ erg cm$^{-2}$ s$^{-1}$,
respectively. We have used only EPIC-pn data in this work. We restrict
the X-ray spectral analysis to sources with more than 70 counts in the
[0.5-7] keV band after background subtraction, for which a reasonable
parameter fit can be obtained. This defines a sample of 98 sources, of
which 76 within an off-axis angle of $10\arcmin$.  The full sample
includes 38 Type-1 AGN, 15 Type-2 AGN, 34 unidentified sources (mostly
newly detected XMM-Newton sources), four extended sources, two normal
galaxies and five stars.

\subsection{The X-ray source catalogue}

The catalogue of the 98 X-ray sources studied here is given in Table
2. We report in the first two columns the source number and the ROSAT
number (if any); in the third column the classification scheme (see
Schmidt et al. \cite{schmidt98}): 1=Type-1 AGN, 2=Type-2 AGN,
3=galaxy, 4=group/cluster of galaxies, 5=star, 9=unidentified source;
in columns 4 and 5 the X-ray source coordinates (J2000).  Off-axis
angles and observed counts in the [0.5-7] keV band are reported in
columns 6 and 7. The X-ray flux is given in three different bands :
[0.5-2], [2-10] and [5-10] keV (see Table 2 in Paper I for the energy
conversion factors in the different bands). Cols. 11, 12 and 13 give
the R and K$^\prime$ magnitudes, and R$-$K$^\prime$ colour
respectively. In column 14, we give the redshift based on
low-resolution Keck spectra. The column density, ${\rm log(N_H)}$ (in
excess to the galactic ones), and the spectral index $\Gamma$ as
measured from spectral fitting are reported in columns 15 and 16. The
errors correspond to $90\%$ confidence level for one interesting
parameter ($\Delta \chi^2=2.706$). The last two columns give the
absorbed X-ray luminosities, which are derived from the X-ray spectra
in the rest-frame bands: [0.5-2] and [2-10] keV. We assume a critical
density universe with ${\rm H_0}=50$ km ${\rm s}^{-1}$ ${\rm
Mpc}^{-1}$ and $\Lambda=0$.

\section{Spectral analysis}

\subsection{Spectra extraction}

The purpose of this work is to perform an X-ray spectral analysis of
the sources in the Lockman Hole field, taking advantage of the large
collecting area of the XMM-Newton satellite.  This represent a step
forward respect to the hardness ratios diagnostic diagrams and stacking
techniques (Tozzi et al. \cite{paolo01}; Alexander et
al. \cite{alexander01a}) in which the range of source redshifts will
smear out the signature of absorption and other X-ray spectral
features (e.g., the iron K$\alpha$ line).

We use an automated procedure to extract the X-ray spectra of the 98
sources. Firstly, a source catalog is constructed using the SAS
detection algorithm (see Paper I for details on the detection
process).  We then perform the source detection using SExtractor
(Bertin et al. $1996$) on the same image ([0.5-7] keV
band). SExtractor yields shape elliptical parameters for each source
(the semi-major/minor axes and the orientation angle) which are added
to the main SAS catalog by cross-correlating the two source lists.

Elliptical parameters for each source are used to define the
appropriate region for the extraction of the spectrum, thus taking
into account the broadening of the PSF at increasing off-axis angles.
The background region is defined as an annulus around the source,
after masking out nearby sources.  The XSELECT tool is used to extract
the spectrum, and the GRPPHA tool is used to bin the data so as to
have at least 20 counts per bin. In this process, the background count
rate is rescaled with the ratio of the source and background areas.

 \begin{figure} \centering
 \includegraphics[width=8cm]{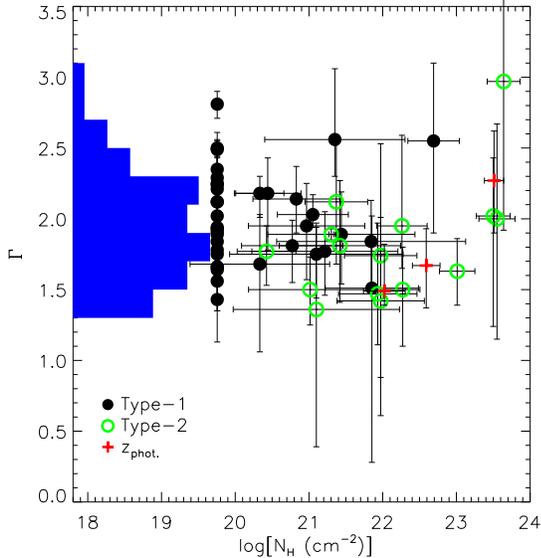} 
   \caption{The power-law photon index ($\Gamma$) versus 
	${\rm log(N_H)}$. Filled 
 circles show the Type-$1$ AGN spectroscopically identified in the
 ROSAT ultradeep HRI survey (Lehmann et al. \cite{ingo01a}) and in the
 on-going optical follow-up of the newly detected XMM-Newton sources
 (PI: Maarten Schmidt), open circles are Type-2
 AGN, crosses are three still unidentified objects for which we have 
photometric redshift estimations (Lehmann et al. \cite{ingo01a}). 
The histogram shows the spectral index distribution (shaded area). 
For both parameters, error bars 
correspond to $90\%$ confidence level for one interesting parameter 
($\Delta \chi^2=2.706$). }
  \label{fig1} 
 \end{figure}

 \begin{figure} \centering 
\includegraphics[width=8cm]{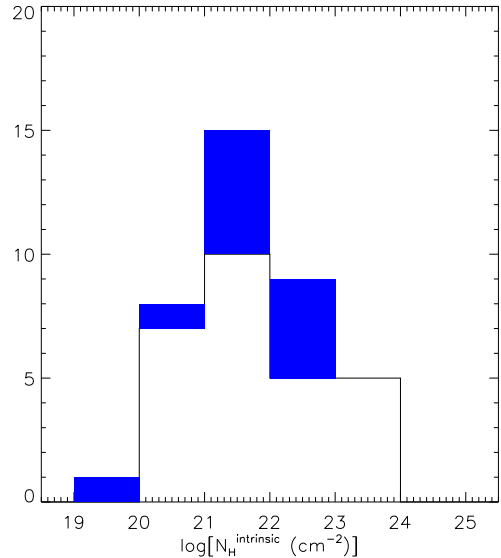}
\caption{Intrinsic ${\rm N_H}$ 
distribution for sources inside an off-axis angle of 10$\arcmin$ with
more than 70 counts in the [0.5-7] keV band. The shaded part indicates
the sources without redshift, for which the measured ${\rm N_H}$ value 
represents a lower limit.
}
\label{fig10}
\end{figure}

\subsection{Spectral parameters}
\label{sec:param}

We use XSPEC (v11.1) for the spectral fitting analysis.  As a first
approximation, a $powerlaw$ model with an intrinsic absorption ($wabs$
or  $zwabs$ if the redshift is known) is used. An additional
photoelectric absorption component ($wabs$) fixed to the Galactic column 
density is also included in the model.

This fit yields the power-law photon index $\Gamma$, the intrinsic
column density ${\rm N_H}$, and the X-ray luminosity in the [0.5-2]
and [2-10] keV rest-frame bands. A clear soft excess is present in
several sources (especially the absorbed ones). In order to reproduce
this feature we add two separate components to the baseline model
({\it wabs$\ast$zwabs$\ast$powerlaw}): a blackbody or an extra
powerlaw. Extra parameters measured from this composite fit (second
power-law index or blackbody temperature) are not reported in Table
2\footnote{In Sect. 7, we present six spectra and best fit models
representative of different classes of objects.}.

\begin{figure*}
\parbox{16.0cm}{\resizebox{\hsize}{!}{\includegraphics{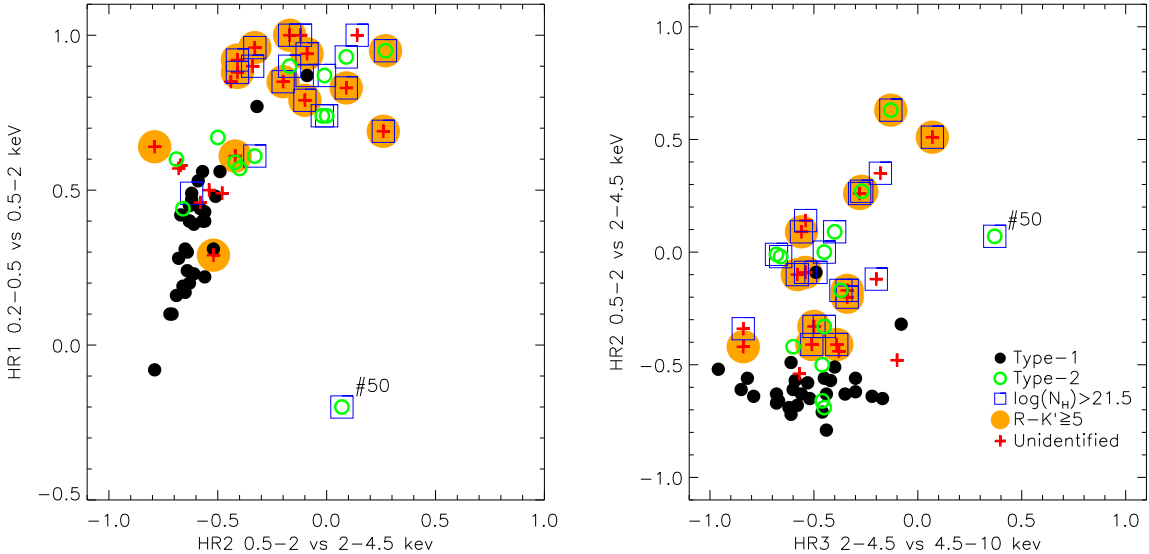}}}
\caption{X-ray diagnostic diagrams based on hardness ratios. Filled 
circles show the Type-$1$ AGN spectroscopically identified in the
ROSAT ultradeep HRI survey (Lehmann et al. \cite{ingo01a}) and in the
on-going optical follow-up of the newly detected XMM-Newton sources 
(PI: Maarten Schmidt). Type-$2$ AGN are marked with open
circles and unidentified sources with crosses. The large filled
circles are EROs (R$-$K$^\prime \geq5$). The sources inside a box are
those with an intrinsic ${\rm log(N_H)} > 21.5$. For clarity, only
sources with hardness ratio errors less than 0.1 are plotted. 
The source $\#50$ is indicated, see Sect. \ref{sec:901} for more details. }
\label{fig2} 
\end{figure*}

In Fig. \ref{fig1}, we plot $\Gamma$ versus the 
column density ${\rm N_H}$ for the sources with known redshifts. 
This diagram suggests that the intrinsic slope of the X-ray
spectrum is the same for all the objects whatever the absorption
levels, with $\langle\Gamma\rangle \simeq 2$. Therefore the increasing
hardness of the source spectra at fainter fluxes observed in the Chandra deep
fields (Giacconi et al. \cite{giacconi01}; Tozzi et
al. \cite{paolo01}; Brandt et al. \cite{brandt01}) is probably due to
intrinsic absorption and not to an intrinsically hard power law.  Several
teams (Della Ceca et al. \cite{rdc99b}; Akiyama et
al. \cite{akiyama00}; Fiore et al. \cite{fiore01}; Maiolino et
al. \cite{maiolino01a}; Page et al. \cite{page01}; Reeves et
al. \cite{reeves00}; Risaliti et al. \cite{risaliti01}) reported the
existence of AGN optically classified as Type-$1$ but with an indication of 
absorption in their X-ray spectra.  In our sample, there are seven
objects with an intrinsic absorption between $10^{21}$ and $10^{22}$
cm$^{-2}$ which are optically classified as Type-$1$ AGN. Moreover, the
source $\#96$ (ROSAT $\#39$) shows a high intrinsic absorption of
$5^{+5}_{-3} \times 10^{22}$ cm$^{-2}$ while it was optically
classified as an unabsorbed QSO at a redshift of $3.279$. In these
cases the optical classification is de-coupled from the
X-ray classification. This could be due to a gas-to-dust
ratio and/or a chemical composition different from those in Galactic
interstellar gas (Akiyama et al. \cite{akiyama00}; Maiolino et
al. \cite{maiolino01b}).  The three sources with photometric redshifts
have an absorption greater than $10^{21.5}$ cm$^{-2}$ which, 
combined with the R$-$K$^\prime$ colours, suggest 
that they are probably obscured AGN (Lehmann et al. \cite{ingo01a}).

\subsection{Observed ${\rm N_H}$ distribution}
\label{sec:N_H}

The ${\rm N_H}$ distribution and its cosmological evolution are key
ingredients in the XRB synthesis models (Comastri et
al. \cite{comastri95}; Gilli et al. \cite{gilli01}). In Fig.
\ref{fig10} we show the ${\rm N_H}$ distribution for the 38 sources with
an off-axis angle $<10\arcmin$. In this central
region where the exposure time is approximately constant, our
threshold of 70 counts in the [0.5-7] keV band corresponds to a flux of
$1.6 \times 10^{-15}$ erg cm$^{-2}$ s$^{-1}$. The surface density of
sources down to this flux limit, $\sim$1700 deg$^{-2}$, is in very
good agreement with that derived from the logN-logS relation given in  
Paper I.  Therefore our sample can be regarded
complete and the derived N$_{\rm H}$ distribution representative
of the overal AGN population at the afore mentioned flux limit. 

\subsection{X-ray colour-colour diagrams}

A useful method to constrain the nature of X-ray sources, in
particular when the signal-to-noise ratio is not high enough for
spectral analysis, is to use X-ray colour-colour diagrams (e.g. Della
Ceca et al. \cite{rdc99a}; Paper I). In Fig. \ref{fig2} we present
two of these diagrams.  We have used the energy bands : 0.2-0.5 keV
(US), 0.5-2 keV (S), 2-4.5 keV (M) and 4.5-10 keV (H) to define three
different hardness ratios: HR1=(S-US)/(S+US), HR2=(M-S)/(M+S),
HR3=(H-M)/(H+M). We use different symbols to indicate Type-$1$ AGN,
Type-$2$ AGN and unidentified sources. The sources with $\log({\rm
N_H})>21.5$ have special labels (square box); when the redshift is
unknown, the derived column densities are only lower limits. We also
highlight the sources with R$-$K$^\prime\geq 5$, usually called
Extremely Red Objects (see Sect. 4.2 for a discussion of the
properties of this class of objects).

In both diagrams, Type-$1$ AGN are confined in a small region, as
opposed to Type-$2$ AGN and unidentified sources which are spread over
a much broader area with high hardness ratios (see also Paper I).
Moreover, using the additional information on the measured intrinsic
absorption column density, it is now clear that the hardening of
non-Type-1 sources is mainly due to the presence of intrinsic
absorption with $\log({\rm N_H})>21.5$ superimposed on relatively soft
spectra (see Fig. \ref{fig1}), rather than intrinsically hard
spectra (this is also consistent with the fact that Type-1 and
Type-2 AGN span similar range in HR3). These diagrams also suggest
that a large fraction of the unidentified sources (mainly newly
detected XMM-Newton sources) are X-ray obscured AGN.

\begin{figure*}
\parbox{16.0cm}{\resizebox{\hsize}{!}{\includegraphics{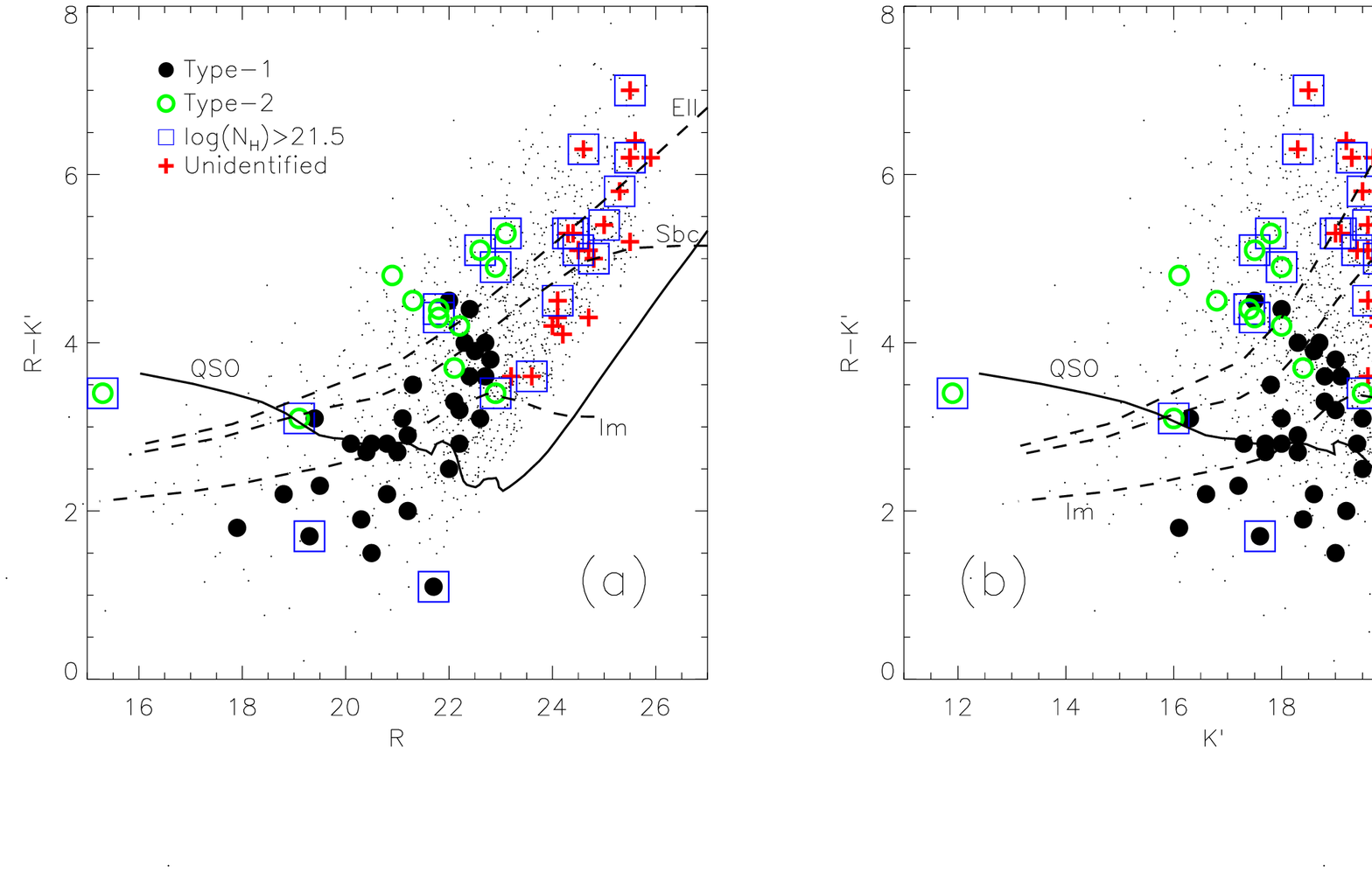}}}
\caption{Color-magnitude diagram, R$-$K$^\prime$ versus R (a) and
R$-$K$^\prime$ versus K$^\prime$ (b), of Lockman Hole sources.
Symbols are as in Fig. \ref{fig2}, while field sources are marked
with small dots. The four evolutionary tracks correspond to an
unreddened QSO with ${\rm L=L^\ast_B}$ and z$=0\div 6$ (solid line),
and to an unreddened elliptical, Sbc and irregular ${\rm L}^\ast$ galaxies
(z$=0\div3$) from the Coleman, Wu \& Weedman (1980) template library
(dashed lines).  }
\label{fig3}
\end{figure*}

\begin{figure} \centering
\includegraphics[width=8cm]{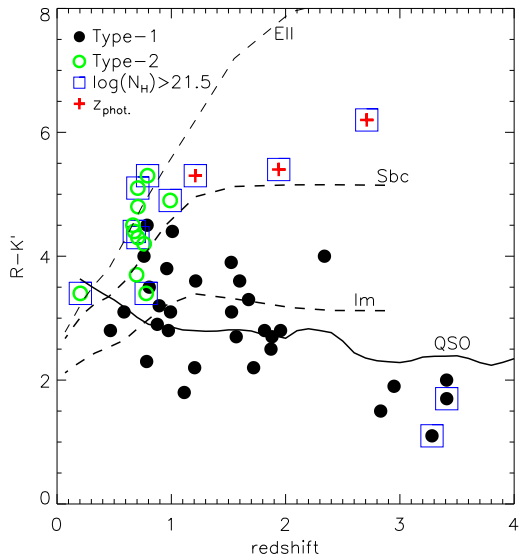} 
\caption{Optical/near-infrared colours as a function of redshift for a sample 
of ROSAT and XMM-Newton sources in the Lockman Hole with optical spectroscopy. 
Symbols are as
in Fig. \ref{fig2}. The evolutionary tracks shown are the same as
in Fig. \ref{fig3}. Type-$1$ AGN have the typical blue colour of a
QSO and are unabsorbed, Type-$2$ AGN follow much redder optical colour
tracks expected for their host galaxy - because the optical nucleus is
obscured - and are intrinsically absorbed.  There are, however,  two 
high-redshift Type-$1$ QSOs whith strong X-ray absorption (see Sect. 
 \ref{sec:qso1}). }
\label{fig4}
\end{figure}

\section{Optical properties}

\subsection{Optical-to-near-IR colours}

A deep broad-band K$^\prime$ (1.92-2.29 $\mu m$) survey of the
Lockman Hole region was carried out with the Omega-Prime camera
(Bizenberger et al. \cite{bizenberger98}) at the Calar Alto 3.5m telescope in
1997 and 1998. This survey covers approximately half the
Lockman Hole field.

Several X-ray surveys (Hasinger et al. \cite{gunther99}; Giacconi et al.
\cite{giacconi01}; Lehmann et al. \cite{ingo01a}; Alexander et al. 
\cite{alexander01a}) have shown that the R$-$K$^\prime$ colour of the optical 
counterparts of X-ray sources 
increases with the optical faintness, and this in a more 
pronounced way than for the field sources (Rosati et al. \cite{piero02}). 
This trend is also evident in the colour-magnitude diagram 
R$-$K$^\prime$ versus R shown in Fig. \ref{fig3}a.
We note that the still unidentified objects are significantly redder
than the bulk of the identified sources.  Using the X-ray information
on ${\rm N_H}$, we also find that there is a strong correlation
between the X-ray absorption and optical colour. For comparison, we also
plot the evolutionary tracks of an early, late and irregular galaxy
type using the template library of Coleman et al. (\cite{coleman80}),
whose spectral energy distributions (SEDs) were extended to the
near-IR and far UV using Bruzual \& Charlot (1993) models as updated
in 2000 (private communication). Magnitudes are normalized to the
measured local value K$^\ast=10.8$, and no dust extinction is
assumed. The QSO evolutionary track is derived from the empirical
template from the Sloan Digital Sky Survey (Vanden Berk et
al. \cite{vanden01}), together with the models of Granato et
al. (\cite{granato97}), normalized to M$_{\rm B}^\ast=-22.4$ (for
brighter objects the curve should be shifted to the left), for the
extension in the near IR. 

In the R$-$K$^\prime$ versus K$^\prime$ diagram, shown in Fig.
\ref{fig3}b, there is no evident trend between R$-$K$^\prime$ colour
and near-IR flux. Moreover, the range of K$^\prime$ magnitudes covered
by the Type-2 AGN and unidentified sources is almost the same as that
of the Type-1 AGN population.  
This is likely due to a combination of a less pronounced absorption 
effect in the K$^\prime$ band, a different K-correction for AGN-type 
spectra (small) and star-like galaxy spectra (large), as well as an 
increased contribution of the host galaxy light in the K$^\prime$ band 
relative to that of the AGN. 
Consequently the difference in the observed magnitudes between absorbed 
and unabsorbed sources is smaller than in the R band.

The R$-$K$^\prime$ versus redshift diagram is shown in Fig.
\ref{fig4} for the subsample with optical identification (redshift and
AGN type) and X-ray spectral fit (see also Fig. 7 in Lehmann et
al. 2001a).  The correlation between optical classification,
optical/near IR colour and X-ray absorption is even clearer than in
Fig. \ref{fig3}.  Most of Type-2 AGN, whose optical colours are
dominated by the host galaxy, are also significantly absorbed (log
N${\rm _H} >21.5$), whereas Type-1 AGN are unobscured and the emission
from the central AGN is contributing significantly to their optical 
colours. There are two exceptions, high redshift Type-1 QSOs, which
are optically unobscured but absorbed in the X-ray band (see
Sect. \ref{sec:qso1}): this could indicate a variation in the
gas-to-dust ratio (Granato et al. \cite{granato97}; Maiolino et
al. \cite{maiolino01a}; Maiolino et al. \cite{maiolino01b}). The
colours of the three sources with photometric redshifts appear to be
dominated by the light from their host galaxies.


The spectroscopic identification is still in progress and, to date, we
have 24 new XMM-Newton sources with measured redshift using LRIS at
the Keck II telescope in March 2001 (PI: Maarten Schmidt). There is an
increasing fraction of Type-2 AGN among these fainter X-ray sources,
and almost all the identified Type-2 AGN are at $z<1$. The
derived but still preliminary redshift distribution seems to be in
clear disagreement with predictions from X-ray background models
(e.g. Gilli et al.
\cite{gilli01})  based on the integrated emission of Type-1 and Type-2 AGN and 
constrained by deep ROSAT surveys (see also Hasinger \cite{gunther02}; Rosati et al. \cite{piero02}).
This calls for a revision of the evolutionary parameters of these
models for both the space density of Type-1 and Type-2 AGN and the
obscuration fraction (Type-1/Type-2 ratio) as a function of the
redshift. The latter is directly related to assumptions in the unified
AGN scheme.

\begin{table}
\caption[]{X-ray detected EROs}
\begin{center}
\begin{tabular}{crrrr}
\hline
 & Ty-$1^a$  & Ty-$2^b$ & Unid.$^c$   & Abs$^d$ \\
\hline
\noalign{\smallskip}
\hline ${\rm R-K}^\prime \geq 3$ & $12$ ($24\%$) & $14$ ($28\%$) & $21$ ($43\%$) & $20$ ($41\%$) \\
\hline ${\rm R-K}^\prime \geq 4$ & $2$ ( $6\%$) & $10$ ($30\%$) & $19$ ($57\%$) & $16$ ($48\%$) \\
\hline ${\rm R-K}^\prime \geq 5$ & $0$ ( $0\%$) & $2$ ($11\%$) & $14$ ($78\%$) & $12$ ($67\%$) \\
\hline ${\rm R-K}^\prime \geq 6$ & $0$ ( $0\%$) & $0$ ($0\%$) & $6$ ($100\%$) & $4$ ($67\%$) \\
\noalign{\smallskip}
\hline
\end{tabular}
\end{center}
$^a$ Type-$1$ AGN \\
$^b$ Type-$2$ AGN \\
$^c$ Unidentified sources \\
$^d$ Sources with log(N$_{\rm H})>21.5$ \\
\label{tab:eros}
\end{table}

\subsection{ X-ray detected Extremely Red Objects} 

In recent years, much efforts have been devoted to understand the
nature of Extremely Red Objects (EROs hereafter). We define EROs as
objects with ${\rm R-K}^{\prime} \geq 5$\footnote{Other selection criteria
have also been used, such as ${\rm R-K} \geq 5.3$ or ${\rm I-K} \geq
4$.}. 

In a recent wide-area survey, Cimatti et al. (\cite{cimatti02}) have
spectroscopically identified a sizeble sample of field EROs with
K$<19.2$ and found them to be almost equally divided between old
passively evolving ellipticals and dusty star-forming galaxies at
$0.7< z < 1.5$.  With XMM-Newton and Chandra observations (Alexander
et al. \cite{alexander01b}; Brusa et al. \cite{brusa02}), the
fraction of optical counterparts with extremely red colours has
significantly increased when compared to the first examples of EROs
found in ROSAT surveys (Lehmann et al. \cite{ingo01a}).

\begin{figure*}
\parbox{16.0cm}{\resizebox{\hsize}{!}{\includegraphics{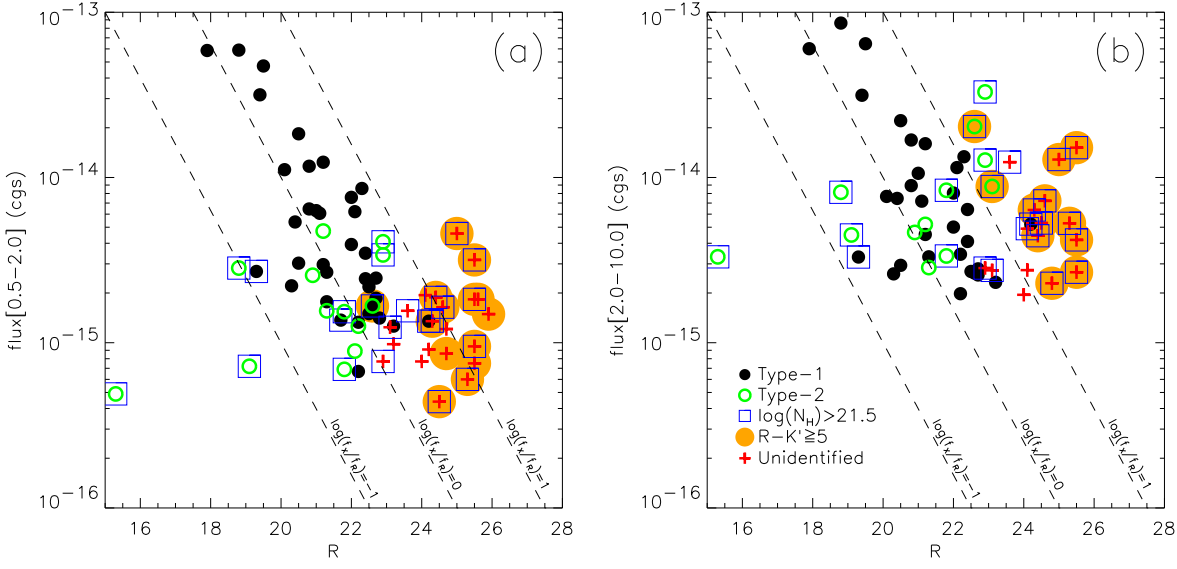}}}
\caption{X-ray flux in the [0.5-2.0] (a) and [2.0-10.] (b)  keV 
bands versus optical R 
magnitudes for those sources in the Lockman Hole with available R band
photometry.  Symbols are as in Fig. \ref{fig2}. The dashed lines are
X-ray-to-optical flux ratio $\log (\frac{f_X}{f_R})$ of $-1$, $0$
and $1$ }
\label{fig5}
\end{figure*}

\begin{figure*}
\parbox{16.0cm}{\resizebox{\hsize}{!}{\includegraphics{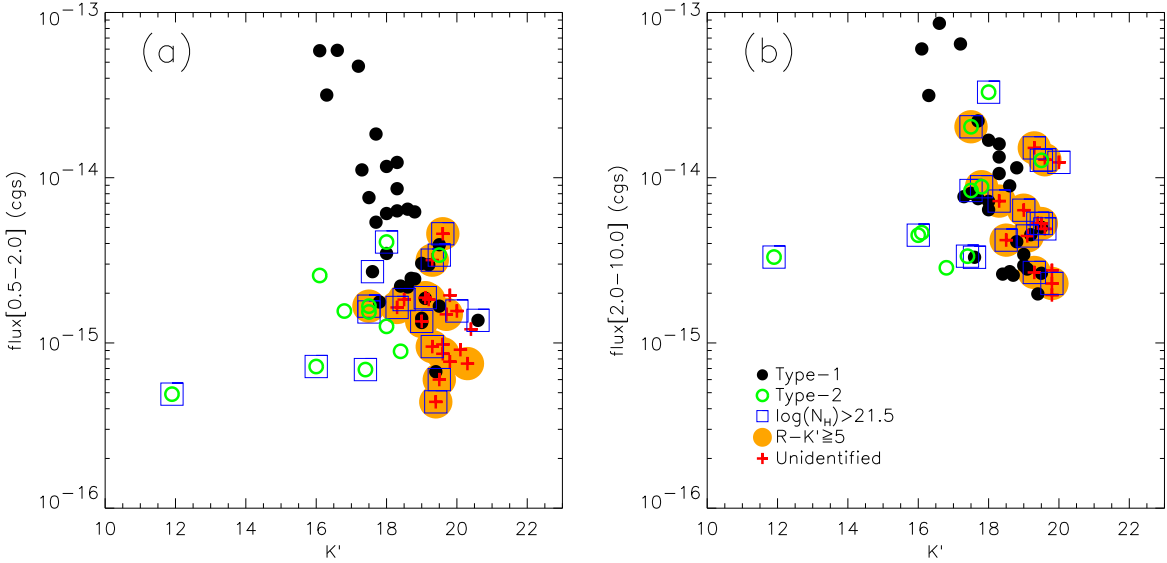}}}
\caption{X-ray flux in the [0.5-2.0] (a) and [2.0-10.] (b) keV bands versus 
optical K$^\prime$ magnitudes for those sources in the Lockman Hole with 
available K$^\prime$ band photometry. Symbols are as in Fig. \ref{fig2}.}
\label{fig9}
\end{figure*}

\begin{figure*}
\parbox{16.0cm}{\resizebox{\hsize}{!}{\includegraphics{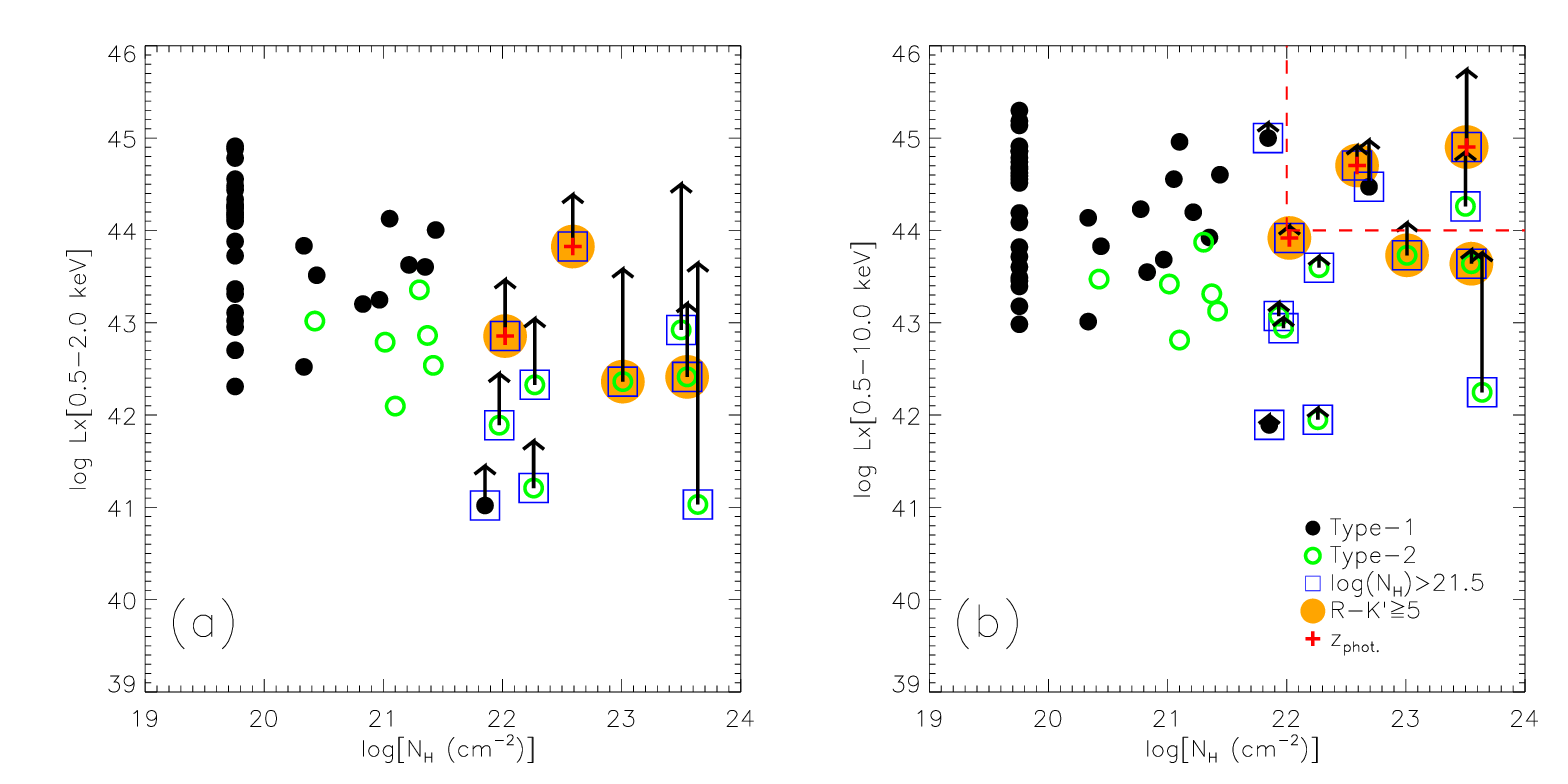}}}
\caption{X-ray luminosity in the [0.5-2] (a) and [0.5-10] (b) keV  
rest-frame band versus the logarithm of the column density ${\rm N_H}$.  Symbols
are as in Fig. \ref{fig2}. Arrows show luminosities corrected for intrinsic 
absorption. The dashed lines in panel b define the ``Type-2 QSO region''. }
\label{fig6}
\end{figure*}

In the subsample of 66 X-ray sources with measured ${\rm
R-K^{\prime}}$ colour, we find 18 (or 27$\%$) EROs. Two of them are
Type-2 AGN, one is classified as a normal galaxy, one is an extended
source and 14 are unidentified sources; no Type-1 AGN are found. From
Table
\ref{tab:eros}, we infer that the fraction of Type-1 AGN decreases
with increasing values of ${\rm R-K^{\prime}}$, whereas the fraction
of unidentified sources and intrinsically absorbed (log(N${\rm
_H})>21.5$) sources increases.  Moreover, all the X-ray detected EROs
have an X-ray-to-optical flux ratio $\log (\frac{{\rm f_X}[2-10 {\rm
keV}]}{{\rm f_R}})>1$ (see Fig. \ref{fig5}b) and they sample the
hardest part of X-ray colour-colour diagrams (see Fig.
\ref{fig2}). The X-ray luminosities in the [0.5-10] keV rest-frame
energy band of the seven EROs with known 
spectroscopic and/or photometric 
redshift are in the range $2.6
\times 10^{42}-8 \times 10^{44}$ erg s$^{-1}$.  We thus conclude that
our X-ray selected sample of EROs is heavily dominated by sources with
strong AGN activity and absorbed X-ray spectra (twelve, or 67$\%$, have
log(N$_{\rm H})>21.5$).

In the 1 Msec observation of the Chandra Deep Field South (Tozzi et
al. \cite{paolo01}; Rosati et al. \cite{piero02}) about 5$\%$ of the
optically selected EROs are detected at X-ray energies, and their
stacked spectrum is consistent with absorbed objects. In that field
19$\%$ of the X-ray sources are EROs, down only to the flux limits of
our complete sample in the Lockman Hole.

\section{X-ray-to-optical flux ratios}

X-ray-to-optical flux ratios can yield important information on the
nature of X-ray sources (Maccacaro et al. \cite{maccacaro88}).  A
value of $-1<\log (\frac{f_X}{f_R}) <1$ is a clear sign of AGN
activity since normal galaxies and stars have usually lower
X-ray-to-optical flux ratios, $\log (\frac{f_X}{f_R}) <-2$. In Fig.
\ref{fig5}, we plot the X-ray flux in [0.5-2.0] (a) and [2.0-10] (b)
keV bands as a function of the R magnitude for the 98 sources of the 
sample. A large fraction of the sources spans the typical X-ray-to-optical
flux ratio of AGN.  While in the soft band (Fig. \ref{fig5}a) the
Type-2 AGN and the unidentified sources are confined at the lower
fluxes of our sample, in the hard band (Fig. \ref{fig5}b), where the
effect of the absorption is weaker, the range of fluxes covered by
these sources is almost the same as that of the Type-1 AGN
sample. 
We also note that $32 \%$ of the sources are confined in a region with 
$\log (\frac{f_X[2-10]}{f_R}) >1$. Among the sources with a 
high X-ray-to-optical flux ratio, $\sim 85\%$ are heavily absorbed 
(log(N$_{\rm H})>21.5$) and $\sim 60 \%$ are EROs.
Their optical classification is still largely incomplete due 
to their faintness: two are Type-1 AGN, four are Type-2 AGN, one is 
an extended source and 13 are unidentified.  

At the current flux limit of our complete sample, the population of
objects with very low X-ray-to-optical flux ratio is largely missing.
Such a population was unveiled by the Chandra deep surveys (Giacconi
et al. \cite{giacconi01}; Hornschemeier et al. \cite{hornschemeier01})
and found to comprise normal galaxies and low-luminosity AGN (with
L$_{\rm X} < 10^{42}$ erg s$^{-1}$ in the [0.5-10] keV energy band).

\begin{figure*}
\parbox{16.0cm}{\resizebox{\hsize}{!}{\includegraphics{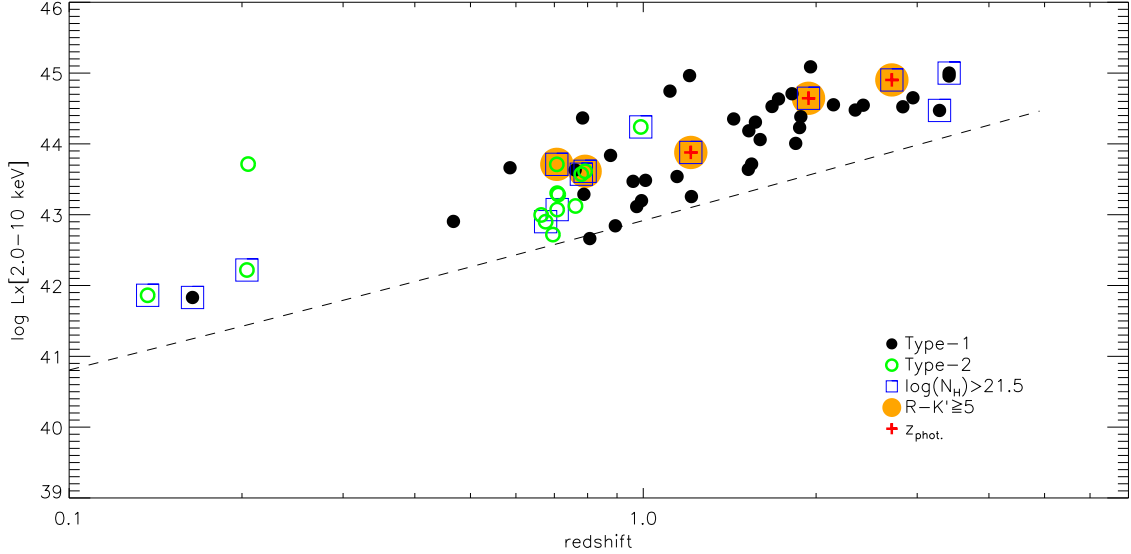}}}
\caption{X-ray luminosity in the [2-10] keV rest-frame band versus redshift. 
Symbols are as in Fig. 2. The dashed line shows the current limit in
the X-ray flux of our sample.}
\label{fig7}
\end{figure*}

There are examples in the literature of ``X-ray bright'' objects
(L$_{\rm X}$[2-10] $>10^{41}$ erg s$^{-1}$) but without any obvious
signature of nuclear activity in the optical spectra (Griffiths
et al. \cite{griffiths95}; Comastri et al. \cite{comastri02} and
references therein). A heavily obscured AGN is among the most likely
explanations. In our sample there are two sources ($\#60$ and $\#92$)
optically classified as normal galaxies (their optical spectra
show emission lines that declare them as star-forming galaxies) which
are however X-ray luminous: L$_{\rm X}$[2-10]$=3.7 \times
10^{41}$ and L$_{\rm X}$[2-10]$=1.9 \times 10^{42}$ erg s$^{-1}$
respectively.  Their X-ray spectra are clearly absorbed (${\rm
N_H}=2^{+2.8}_{-1.6}\times10^{23}$ and ${\rm
N_H}=9^{+3.2}_{-1.4}\times10^{20}$ ${\rm cm}^{-2}$) reinforcing the
evidence that they contain an obscured AGN. More examples of this
class of objects are expected at the completion of the optical
identification of the newly detected XMM-Newton sources.

In Fig. \ref{fig9}, we plot the X-ray flux in [0.5-2.0] (a) and
[2.0-10] (b) keV bands as a function of K$^\prime$ magnitudes.  In
these diagrams, there is a strong overlap between Type-2 AGN and
unidentified sources and the Type-1 AGN population, more prounonced in
Fig. \ref{fig9}b. The overlap in magnitudes is likely due 
mainly to a K-correction effect (see also Fig. \ref{fig3}b), 
whereas in the X-ray hard band the effect of the absorption is weaker 
(see also Fig. \ref{fig5}b).

Finally, in the soft band (Fig. \ref{fig5}a and \ref{fig9}a) the
fraction of absorbed objects increases significantly as the flux
decreases. This inevitably leads, in a flux limited sample, to a bias 
in the N$_{\rm H}$ distribution at high value of N$_{\rm H}$ (see Fig.
\ref{fig10}).

\section{X-ray luminosity and Type-2 QSO candidates}

We have redshifts (and luminosities) for 61 objects or $\sim62\%$ of
the sample with X-ray spectral analysis.  In Fig. \ref{fig6}, we
plot X-ray luminosity in the [0.5-2] (a) and [0.5-10] (b) keV
rest-frame bands as a function of the log(${\rm N_H}$).  Type-1 AGN
(objects without soft absorption) cover a range between $1 \times
10^{41}$ erg s$^{-1}$ and $9 \times 10^{44}$ erg s$^{-1}$ in the
[0.5-2] keV band; whereas absorbed Type-2 AGN have luminosities in the
range $1 \times 10^{41}$ - $2 \times 10^{43}$ erg s$^{-1}$. In the
total band (Fig. \ref{fig6}b) the effect of absorption is less
evident so that the range of luminosity of Type-1 ($1\times 10^{42}$ -
$ 2\times 10^{45}$ erg s$^{-1}$) and Type-2 AGN ($1\times 10^{42}$ -
$2\times 10^{44}$ erg s$^{-1}$) is comparable. We have derived the
unabsorbed luminosities for objects with log(N${\rm_H)>21.5}$ and
reported them in Fig. \ref{fig6} as arrows. In the soft band (Fig.
\ref{fig6}a), where the effect of absorption is stronger, 
luminosities increase substantially and the range of intrinsic
luminosities of Type-2 AGN fall in the same range as that of Type-1's
(see also Gilli et al., in preparation).  In Fig. \ref{fig6}b, we
have highlighted the region where L$_{\rm X}$[0.5-10]$>10^{44}$ erg
s$^{-1}$ and ${\rm log(N_H)}>22$ cm$^{-2}$, i.e. the ``Type-2 QSO 
region''.  Six objects fall inside this area: one is optically
classified as a Type-1 AGN (see Sect. \ref{sec:qso1} for more
details), two are Type-2 AGN.  For the remaining three, we derived
photometric redshifts and due to their X-ray absorption and
optical/near-IR colours are likely Type-2 AGN.  Four of them are also
EROs. We argue that these six sources are reliable Type-2 QSO
candidates.  All of them are within an off-axis angle of 10$\arcmin$
where the sample is complete (see Sect. \ref{sec:N_H}) and we thus
derive a density of $\sim$69 objects of this class per square degree.

In Fig. \ref{fig7}, we show the X-ray luminosity as a function of
redshift, using the observed hard band luminosity which is relatively
unaffected by absorption.

\section{High S/N spectra}

\begin{figure*}
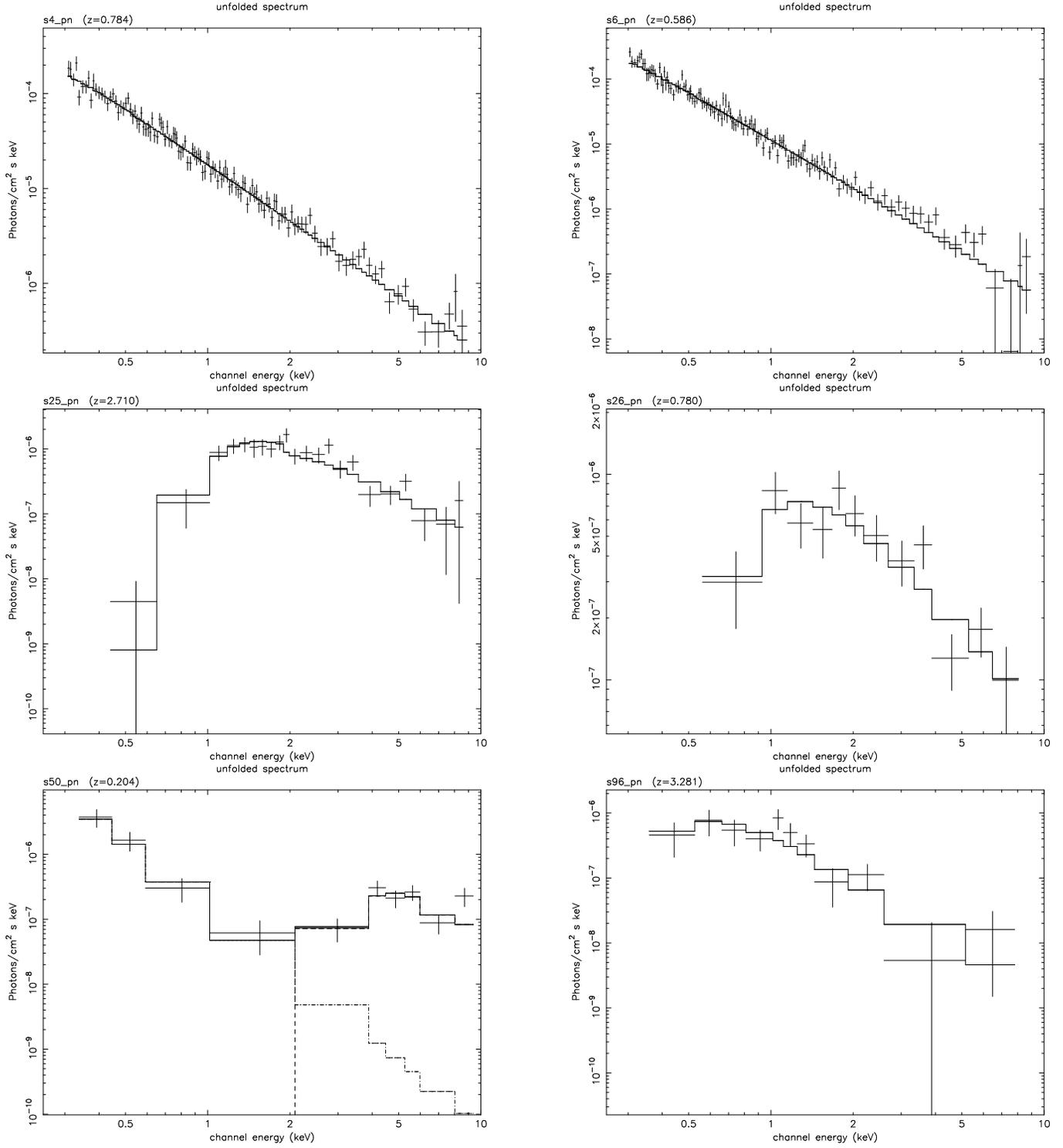

\parbox{8.3cm}{\resizebox{\hsize}{!}{\includegraphics[angle=270]{4.ps}}}
\hfill
\parbox{8.3cm}{\resizebox{\hsize}{!}{\includegraphics[angle=270]{6.ps}}}
\\
\parbox{8.3cm}{\resizebox{\hsize}{!}{\includegraphics[angle=270]{25.ps}}}
\hfill
\parbox{8.3cm}{\resizebox{\hsize}{!}{\includegraphics[angle=270]{26.ps}}}
\\
\parbox{8.3cm}{\resizebox{\hsize}{!}{\includegraphics[angle=270]{50.ps}}}
\hfill
\parbox{8.3cm}{\resizebox{\hsize}{!}{\includegraphics[angle=270]{96.ps}}} \\
\caption{XMM EPIC-pn CCD spectra and best fit models. Top: source $\#4$ 
(ROSAT $\#29$) and source $\#6$ (ROSAT $\#16$); middle: source $\#25$ 
(ROSAT $\#84$) and source $\#26$ (ROSAT $\#117$); bottom: source $\#50$ 
(ROSAT $\#901$) and source $\#96$ (ROSAT $\#39$).}
\label{fig8}
\end{figure*}

In Fig. \ref{fig8} we show six X-ray spectra representative of the
different classes of objects in our sample. The source numbers refer
to the catalogue presented in this work, for reference we give also
the ROSAT catalogue numbers (Hasinger et al., \cite{gunther98}).  The
redshift of the sources are reported in Lehmann et
al. (\cite{ingo01a}).  Sources with interesting line features will be
reported in a future work (Hasinger et al. 2002, in preparation).

\subsection{Unabsorbed sources}

Source $\#4$ (ROSAT $\#29$): this source was already observed by ROSAT
(Lehmann et al. \cite{ingo00}) and optically classified as a Type-1
AGN at $z=0.784$.
This is one of the brightest sources in our sample (3164 EPIC-pn
counts in the [0.5-7] keV band).
This source is very well fitted ($\chi_\nu^2=1.03$) by a simple power
law model with $\Gamma=2.02^{+0.04}_{-0.04}$ and ${\rm N_H}$
consistent with the Galactic value ($5.7\times10^{19}$ ${\rm
cm}^{-2}$).  We measure ${\rm L_X} = 4.2\times10^{44}$ erg s$^{-1}$ in
the [0.5-10] keV rest-frame band, and ${\rm log
(\frac{f_x}{f_R})}=0.6$. \\ Source $\#6$ (ROSAT $\#16$): it was
observed by ROSAT (Schmidt et al.\cite{schmidt98}) and classified as a
Type-1 AGN at $z=0.586$.
This source (1537 EPIC-pn counts in the [0.5-7] keV band) is well
fitted ($\chi_\nu^2=1.15$) by a simple power law model with
$\Gamma=2.47_{-0.03}^{+0.08}$ and N$_{\rm H}$ consistent with the
Galactic value ($5.7\times10^{19}$ ${\rm cm}^{-2}$). It has L$_{\rm X}
= 1.2
\times10^{44}$ erg s$^{-1}$ in the [0.5-10] keV rest-frame band.\\

\subsection{Absorbed sources}

Source $\#25$ (ROSAT $\#84$): this object was part of the ROSAT
ultradeep HRI survey (Hasinger et al., \cite{gunther98}). Lehmann et
al. (\cite{ingo01a}) give a photometric redshift $z_{\rm
phot}=2.71^{+0.29}_{-0.41}$.
The spectrum extracted from the EPIC-pn data (332 counts in the
[0.5-7] keV band) is well fitted ($\chi_\nu^2=1.09$) by a $wabs\ast
zwabs(powerlaw)$ model, with an intrinsic absorption of ${\rm
N_H}=3^{+1}_{-1}\times10^{23}$ ${\rm cm}^{-2}$ and
$\Gamma=2.3^{+0.3}_{-0.3}$; the unabsorbed rest-frame luminosity in
the [0.5-10] keV band is ${\rm L_X}=5.49 \times 10^{45}$ erg
s$^{-1}$.\\ Source $\#26$ (ROSAT $\#117$): was observed by ROSAT
(Schmidt et al. \cite{schmidt98}) and optically classified as a Type-$2$ AGN at
$z=0.780$. From the fit of the X-ray spectra we get
the values, ${\rm N_H}=2^{+1}_{-1}\times10^{22}$ cm$^{-2}$ and
$\Gamma=1.5^{+0.4}_{-0.3}$. The unabsorbed X-ray luminosity in the
[0.5-10] keV rest-frame band is ${\rm L_X}=4 \times 10^{43}$
erg~s$^{-1}$.\\


\subsection{Multi component spectra}
\label{sec:901}

Source $\#50$ (ROSAT $\#901$): this source was classified as a Type-2
AGN at z=0.204 by Lehmann et al. (\cite{ingo01a}).  
As noted in Paper I, a very soft component superimposed on a heavy
absorbed power law, is likely present in this source as suggested by 
the unusually large value of the hardness ratio HR3.  The XMM-Newton
spectrum clearly shows such a feature. By fitting a double power law
model ($wabs(zwabs(powerlaw)+powerlaw)$), we obtain: ${\rm
N_H}=4^{+2.5}_{-1.5}\times 10^{23}$ cm$^{-2}$,
$\Gamma=3^{+1}_{-1}$ for the hard component and
$\Gamma=3.3^{+0.4}_{-0.5}$ for the soft component ($\chi_\nu^2=1.2$).
We also find an unabsorbed X-ray luminosty L$_{\rm X}=5.7 \times
10^{43}$ erg s$^{-1}$ in the [0.5-10] keV rest-frame band and a ratio
${\rm log (\frac{f_x}{f_R})}=-2.4$, unusually low for an AGN, which is
probably due to the strong intrinsic absorption.

\subsection{Type-1 QSO with X-ray absorption}
\label{sec:qso1}

Source $\#96$ (ROSAT $\#39$): this object is optically classified as
a Type-1 QSO at  $z=3.279$ (Lehmann et al., \cite{ingo01a}).
A clear absorption is present in the X-ray spectrum 
and the fit yields ${\rm N_H}=5^{+5}_{-3}\times
10^{22}$ cm$^{-2}$.  As already argued in Sect. \ref{sec:param}, this 
mismatch between the optical and X-ray classifications could be due to a
gas-to-dust ratio or a chemical composition different from that of 
the Galactic interstellar gas (Akiyama et al. \cite{akiyama00}; Maiolino 
et al. \cite{maiolino01b}).

\section{Conclusions}

We have discussed the X-ray spectral properties of a sample 
of 98 sources found in the 100 ksec XMM-Newton observation of the Lockman
Hole, using data from the EPIC-pn
detector. The large throughput and the unprecedented sensitivity at
high energies of the X-ray telescope and detectors allow us, for the
first time, to measure separetely the intrinsic absorption and the
slope of the power law emission spectrum for the faint source population.  
We have derived the spectral index ($\Gamma$) and the column density
(${\rm N_H}$) for sources with more than 70 counts in the [0.5-7] keV 
band. We find that the value of $\Gamma$ is independent of
the absorption level 
with $<\Gamma> \approx 2$. Thus, we infer that
the progressive hardening of the X-ray spectra of faint sources
observed in Chandra deep fields (Giacconi et al. \cite{giacconi01}; 
Tozzi et al. \cite{paolo01}; Brandt et al. \cite{brandt01}) is 
mainly due to the increasing level of intrinsic absorption rather 
than intrinsically flat spectra. 

We confirm that the ${\rm R-K}^\prime$ colours of X-ray counterparts
get redder towards fainter R magnitudes. Such a trend is not present
between ${\rm R-K}^\prime$ and the K$^\prime$ magnitude; this is
likely due to a combination of a less pronounced absorption effect in
this band, a different K-correction for AGN-type spectra (small) and
star-like galaxy spectra (large), as well as an increased contribution
of the host galaxy light in the K$^\prime$ band relative to that of
the AGN.

Comparing the ${\rm R-K}^\prime$ colours of the X-ray sources with
evolutionary tracks of various galaxy-types as a function of
redshift, we find that Type-2 AGN have colours dominated by the host
galaxy and are also significantly absorbed (log ${\rm N_H}>21.5$). On
the other hand, for Type-1 AGN, the large majority of which are
unabsorbed, the nuclear component is significantly contributing to
their optical colours.  In addition, there is a strong correlation between the
${\rm R-K}^\prime$ colour and the amount of intrinsic X-ray
absorption.

We have also defined an X-ray selected sample of 18 EROs (${\rm
R-K}^\prime \geq 5$) and found that it mainly comprises X-ray absorbed
objects with a strong correlation between colour and 
intrinsic column density.

We have derived the unabsorbed rest-frame luminosities of the sources
with strong intrinsic absorption. There are six absorbed, bright X-ray
objects in our sample with ${\rm L_X}[0.5-10]>10^{44}$ erg s$^{-1}$
and ${\rm log(N_H)}>10^{22}$ cm$^{-2}$: one is an optically classified
Type-1 QSO (source $\#96$ see Sect. \ref{sec:qso1}), two are Type-2
AGN and the remaining three have a photometric redshift and due to
their X-ray absorption and optical/near-IR colours likely Type-2 AGN.
Four of them are also EROs (${\rm R-K}^\prime \geq 5$). These are
likely to be Type-2 QSO candidates and we derive a density of 
$\sim 69$ objects of this class per square degree at a flux limit 
in the [0.5-7] keV band of $1.6 \times 10^{-15}$ erg cm$^{-2}$ s$^{-1}$.

Our analysis of the unidentified sources (mostly newly detected XMM
sources) shows that the majority of these sources have absorbed X-ray
spectra and are consequently located in the harder part of the
diagnostic X-ray colour-colour diagrams.
They are also optically fainter ($\sim80\%$ of them have R$>24$) and
their optical-to-near-IR colours are redder ($\sim90\%$ have
R$-$K$^\prime \geq 4$) than already identified sources.
Their X-ray-to-optical flux ratios are $\log (\frac{f_X[2-10]}{f_R})>1$.
From these properties, we argue that the majority of these sources are
Type-2 AGN. This is confirmed by our on-going optical spectroscopic
survey which is showing that the bulk of these sources is at $z<1$.
Two X-ray bright optically ``normal'' galaxies are present in our
sample. Their X-ray spectra are clearly absorbed suggesting the
presence of an obscured AGN. We expect this class of objects to
increase from the optical identification of the newly detected
XMM-Newton sources.

\begin{acknowledgements}
  We thank Andrea Comastri, Roberto Gilli, Giorgio Matt and Paolo
  Tozzi for useful comments and discussions. We thank the referee,
  X. Barcons, for helpful comments that improved the manuscript. RDC
  acknowledge financial support from the Italian Space Agency, ASI
  (I/R/037/01), under the project "Cosmologia Osservativa con
  XMM-Newton".
\end{acknowledgements}

\newpage

\onecolumn
\begin{sidewaystable}[h!]
\centering
\bigskip
{\scriptsize
\begin{tabular}{rrcccrrrrrcccccccc}
\multicolumn{18}{l}{\footnotesize {\bf Table 2.} X-ray catalogue} \\
\smallskip \\
\hline
\vspace{-2 mm} \\
XID   & Rosat & Type$^a$ & RA & Dec & Off-axis & counts &  Flux$^b$ & Flux$^b$ & Flux$^b$ & R & K$^\prime$ & R-K$^\prime$ & z & ${\rm log(N_H)}^c$ & $\Gamma^d$ & L$_X$$^e$ & L$_X$$^e$\\
 &  &  &   &   & angle$^f$ & [0.5-7] & [0.5-2] & [2-10] & [5-10] &  &  &  &  &  &  & [0.5-2] & [2-10]\\
\vspace{-2 mm} \\
\hline
\vspace{-2 mm} \\
1 & 28 & 2 & 10 54 21.3 & +57 25 43 & 13.58 & 6071 & 11.06 & 20.31 & 9.30 & .... & .... & .... & 0.205 & 21.29$_{21.26}^{21.32}$ & 1.89$_{1.85}^{1.93}$ & 43.356 & 43.715\smallskip \\
2 & 32 & 1 & 10 52 39.6 & +57 24 32 & 4.28 & 4440 & 5.87 & 6.02 & 2.97 & 17.9 & 16.1 & 1.8 & 1.113 & 0.00 & 2.49$_{2.45}^{2.53}$ & 44.911 & 44.745\smallskip \\
3 & 6 & 1 & 10 53 16.8 & +57 35 52 & 8.40 & 3590 & 5.90 & 8.60 & 3.86 & 18.8 & 16.6 & 2.2 & 1.204 & 0.00 & 1.91$_{1.87}^{1.94}$ & 44.783 & 44.964\smallskip \\
4 & 29 & 1 & 10 53 35.1 & +57 25 43 & 7.65 & 3164 & 4.73 & 6.45 & 2.62 & 19.5 & 17.2 & 2.3 & 0.784 & 0.00 & 2.02$_{1.98}^{2.06}$ & 44.276 & 44.366\smallskip \\
5 & 8 & 5 & 10 51 30.9 & +57 34 39 & 11.30 & 1958 & 4.66 & 0.65 & 0.00 & .... & .... & .... & ..... & ...... & ...... & ...... & ......\smallskip \\
6 & 16 & 1 & 10 53 39.7 & +57 31 05 & 7.95 & 1537 & 3.16 & 3.15 & 1.33 & 19.4 & 16.3 & 3.1 & 0.586 & 0.00 & 2.50$_{2.45}^{2.56}$ & 43.880 & 43.663\smallskip \\
7 & 0 & 5 & 10 53 00.5 & +57 42 10 & 13.58 & 1271 & 3.58 & 0.00 & 0.00 & .... & .... & .... & ..... & ...... & ...... & ...... & ......\smallskip \\
8 & 31 & 1 & 10 53 31.9 & +57 24 56 & 7.64 & 1208 & 1.84 & 2.21 & 1.16 & 20.5 & 17.7 & 2.8 & 1.956 & 0.00 & 1.93$_{1.86}^{1.99}$ & 44.882 & 45.088\smallskip \\
9 & 232 & 5 & 10 53 36.3 & +57 38 01 & 11.67 & 962 & 2.42 & 0.00 & 0.00 & .... & .... & .... & ..... & ...... & ...... & ...... & ......\smallskip \\
10 & 25 & 1 & 10 53 44.9 & +57 28 41 & 8.33 & 746 & 1.17 & 1.69 & 0.50 & 20.8 & 18.0 & 2.8 & 1.816 & 0.00 & 1.89$_{1.80}^{1.91}$ & 44.484 & 44.708\smallskip \\
11 & 9 & 1 & 10 51 54.3 & +57 34 36 & 8.74 & 744 & 1.24 & 1.60 & 0.74 & 21.2 & 18.3 & 2.9 & 0.877 & 20.20$_{19.76}^{20.30}$ & 2.18$_{2.06}^{2.35}$ & 43.833 & 43.837\smallskip \\
12 & 513 & 1 & 10 52 54.4 & +57 23 42 & 5.32 & 681 & 0.86 & 1.33 & 0.62 & 22.3 & 18.3 & 4.0 & 0.761 & 0.00 & 1.75$_{1.65}^{1.88}$ & 43.365 & 43.631\smallskip \\
13 & 37 & 1 & 10 52 48.1 & +57 21 17 & 7.54 & 612 & 1.11 & 0.77 & 0.34 & 20.1 & 17.3 & 2.8 & 0.467 & 0.00 & 2.81$_{2.72}^{2.91}$ & 43.310 & 42.905\smallskip \\
14 & 2 & 1 & 10 52 30.1 & +57 39 13 & 10.57 & 599 & 1.14 & 1.55 & 0.48 & .... & .... & .... & 1.437 & 21.03$_{20.30}^{21.36}$ & 2.03$_{1.89}^{2.12}$ & 44.127 & 44.352\smallskip \\
15 & 12 & 2 & 10 51 48.6 & +57 32 47 & 8.32 & 570 & 0.41 & 3.29 & 2.17 & 22.9 & 18.0 & 4.9 & 0.990 & 23.50$_{23.29}^{23.74}$ & 2.02$_{1.61}^{2.80}$ & 42.922 & 44.239\smallskip \\
16 & 77 & 1 & 10 52 59.1 & +57 30 29 & 2.75 & 516 & 0.62 & 1.15 & 0.70 & 22.1 & 18.8 & 3.3 & 1.676 & 0.00 & 1.56$_{1.45}^{1.71}$ & 44.100 & 44.528\smallskip \\
17 & 0 & 9 & 10 54 07.1 & +57 35 27 & 13.09 & 511 & 1.35 & 1.44 & 0.63 & .... & .... & .... & ..... & 0.00 & 2.29$_{2.20}^{2.38}$ & ...... & ......\smallskip \\
18 & 426 & 1 & 10 53 03.8 & +57 29 24 & 2.87 & 443 & 0.76 & 0.81 & 0.00 & 22.0 & 17.5 & 4.5 & 0.788 & 20.79$_{19.76}^{20.90}$ & 2.14$_{1.91}^{2.38}$ & 43.202 & 43.288\smallskip \\
19 & 14 & 9 & 10 52 42.2 & +57 31 58 & 3.18 & 427 & 0.46 & 1.28 & 0.64 & 25.0 & 19.6 & 5.4 & 1.94$_{phot.}^g$ & 22.59$_{22.39}^{22.77}$ & 1.67$_{1.41}^{1.97}$ & 43.825 & 44.644\smallskip \\
20 & 120 & 1 & 10 53 09.3 & +57 28 21 & 3.57 & 426 & 0.54 & 0.75 & 0.45 & 20.4 & 17.7 & 2.7 & 1.568 & 0.00 & 2.12$_{1.98}^{2.28}$ & 44.192 & 44.307\smallskip \\
21 & 27 & 1 & 10 53 51.5 & +57 27 04 & 9.37 & 412 & 0.65 & 0.89 & 0.56 & 20.8 & 18.6 & 2.2 & 1.720 & 0.00 & 1.66$_{1.56}^{1.76}$ & 44.258 & 44.634\smallskip \\
22 & 5 & 1 & 10 53 02.5 & +57 37 57 & 9.53 & 387 & 0.63 & 1.06 & 0.48 & 21.0 & 18.3 & 2.7 & 1.881 & 0.00 & 1.94$_{1.79}^{2.11}$ & 44.169 & 44.383\smallskip \\
23 & 30 & 1 & 10 52 57.3 & +57 25 06 & 4.16 & 368 & 0.61 & 0.72 & 0.00 & 21.1 & 18.0 & 3.1 & 1.527 & 0.00 & 2.35$_{2.21}^{2.50}$ & 44.239 & 44.185\smallskip \\
24 & 0 & 2 & 10 52 37.8 & +57 33 22 & 4.62 & 352 & 0.17 & 2.03 & 1.61 & 22.6 & 17.5 & 5.1 & 0.707 & 23.01$_{22.70}^{23.18}$ & 1.63$_{1.40}^{1.87}$ & 42.362 & 43.711\smallskip \\
25 & 84 & 9 & 10 52 17.0 & +57 20 17 & 9.21 & 332 & 0.32 & 1.51 & 0.75 & 25.5 & 19.3 & 6.2 & 2.71$_{phot.}^g$ & 23.51$_{23.37}^{23.63}$ & 2.27$_{1.92}^{2.64}$ & ...... & 44.903\smallskip \\
26 & 117 & 2 & 10 53 48.6 & +57 30 35 & 8.99 & 321 & 0.34 & 1.27 & 1.02 & 22.9 & 19.5 & 3.4 & 0.780 & 22.27$_{22.06}^{22.53}$ & 1.50$_{1.20}^{1.90}$ & 42.327 & 43.572\smallskip \\
27 & 45 & 2 & 10 53 19.2 & +57 18 52 & 11.06 & 287 & 0.47 & 0.52 & 0.00 & 21.2 & .... & .... & 0.711 & 20.32$_{19.76}^{20.48}$ & 1.77$_{1.59}^{2.01}$ & 43.018 & 43.284\smallskip \\
28 & 23 & 1 & 10 52 24.6 & +57 30 10 & 2.83 & 286 & 0.35 & 0.64 & 0.00 & 22.4 & 18.0 & 4.4 & 1.009 & 20.94$_{19.76}^{21.32}$ & 1.95$_{1.65}^{2.33}$ & 43.248 & 43.485\smallskip \\
29 & 13 & 1 & 10 52 13.1 & +57 32 22 & 5.36 & 277 & 0.39 & 0.50 & 0.00 & 22.0 & 19.5 & 2.5 & 1.872 & 20.73$_{19.76}^{21.11}$ & 1.81$_{1.63}^{2.07}$ & ...... & 44.230\smallskip \\
30 & 52 & 1 & 10 52 43.4 & +57 15 45 & 13.04 & 250 & 0.57 & 0.46 & 0.00 & .... & .... & .... & 2.144 & 0.00 & 2.29$_{2.15}^{2.48}$ & 44.555 & 44.552\smallskip \\
31 & 104 & 2 & 10 52 41.3 & +57 36 50 & 8.05 & 224 & 0.28 & 0.81 & 0.49 & 18.8 & .... & .... & 0.137 & 22.26$_{21.60}^{22.29}$ & 1.95$_{1.31}^{2.25}$ & 41.209 & 41.862\smallskip \\
32 & 491 & 9 & 10 52 25.3 & +57 25 50 & 3.79 & 216 & 0.16 & 0.72 & 0.49 & 24.6 & 18.3 & 6.3 & ..... & 22.04$_{21.73}^{22.30}$ & 2.17$_{1.52}^{2.89}$ & ...... & ......\smallskip \\
33 & 123 & 9 & 10 51 28.3 & +57 27 37 & 10.11 & 209 & 0.16 & 1.24 & 0.60 & 23.6 & 20.0 & 3.6 & ..... & 22.15$_{21.83}^{22.40}$ & 1.90$_{1.30}^{2.34}$ & ...... & ......\smallskip \\
34 & 80 & 1 & 10 51 44.6 & +57 28 07 & 7.87 & 207 & 0.30 & 0.45 & 0.00 & 21.2 & 19.2 & 2.0 & 3.409 & 21.08$_{19.76}^{22.11}$ & 1.75$_{1.53}^{2.06}$ & ...... & 44.958\smallskip \\
35 & 116 & 2 & 10 52 37.5 & +57 31 05 & 2.40 & 206 & 0.26 & 0.47 & 0.32 & 20.9 & 16.1 & 4.8 & 0.708 & 20.99$_{19.76}^{21.41}$ & 1.50$_{1.27}^{1.75}$ & 42.789 & 43.304\smallskip \\
36 & 814 & 1 & 10 52 45.2 & +57 21 21 & 7.45 & 199 & 0.30 & 0.29 & 0.00 & 20.5 & 19.0 & 1.5 & 2.832 & 0.00 & 2.12$_{1.88}^{2.32}$ & 44.335 & 44.524\smallskip \\
37 & 486 & 9 & 10 52 43.2 & +57 28 00 & 0.79 & 191 & 0.19 & 0.44 & 0.34 & 24.4 & 19.1 & 5.3 & 1.21$_{phot.}^g$ & 22.02$_{21.72}^{22.29}$ & 1.49$_{1.16}^{1.59}$ & 42.855 & 43.879\smallskip \\
38 & 0 & 2 & 10 52 06.6 & +57 29 24 & 4.92 & 187 & 0.15 & 0.84 & 0.34 & 21.8 & 17.5 & 4.3 & 0.708 & 21.93$_{21.23}^{22.31}$ & 1.47$_{0.97}^{1.83}$ & ...... & 43.072\smallskip \\
39 & 0 & 9 & 10 54 04.5 & +57 20 36 & 13.72 & 179 & 0.44 & 0.00 & 0.00 & .... & .... & .... & ..... & 0.00 & 2.74$_{2.57}^{3.00}$ & ...... & ......\smallskip \\
40 & 20 & 5 & 10 54 10.5 & +57 30 40 & 11.90 & 178 & 0.43 & 0.00 & 0.00 & .... & .... & .... & ..... & ...... & ...... & ...... & ......\smallskip \\
\vspace{-2 mm} \\
\hline
\vspace{-2 mm} \\
\end{tabular}
}
\normalsize
\end{sidewaystable}
\twocolumn

\onecolumn
\begin{sidewaystable}[h!]
\centering
\bigskip
{\scriptsize
\begin{tabular}{rrcccrrrrrcccccccc}
\multicolumn{18}{l}{\footnotesize {\bf Table 2.}  (continued)} \\
\smallskip \\
\hline
\vspace{-2 mm} \\
XID   & Rosat & Type$^a$ & RA & Dec & Off-axis & counts &  Flux$^b$ & Flux$^b$ & Flux$^b$ & R & K$^\prime$ & R-K$^\prime$ & z & ${\rm log(N_H)}^c$ & $\Gamma^d$ & L$_X$$^e$ & L$_X$$^e$\\
 &  &  &  &   & angle$^f$ & [0.5-7] & [0.5-2] & [2-10] & [5-10] &  &  &  &  &  &  & [0.5-2] & [2-10]\\
\vspace{-2 mm} \\
\hline
\vspace{-2 mm} \\
41 & 0 & 2 & 10 53 05.4 & +57 28 10 & 3.08 & 176 & 0.00 & 0.88 & 1.15 & 23.1 & 17.8 & 5.3 & 0.792 & 23.55$_{23.31}^{23.80}$ & 2.00$_{1.33}^{2.85}$ & 42.413 & 43.613\smallskip \\
42 & 821 & 1 & 10 53 22.1 & +57 28 52 & 5.26 & 170 & 0.25 & 0.26 & 0.00 & 22.7 & 18.7 & 4.0 & 2.340 & 21.43$_{19.76}^{21.75}$ & 1.89$_{1.59}^{2.24}$ & 44.004 & 44.478\smallskip \\
43 & 75 & 1 & 10 51 25.3 & +57 30 48 & 10.62 & 161 & 0.27 & 0.33 & 0.57 & 19.3 & 17.6 & 1.7 & 3.410 & 21.84$_{19.76}^{22.32}$ & 1.84$_{1.55}^{2.15}$ & ...... & 45.000\smallskip \\
44 & 477 & 1 & 10 53 05.6 & +57 34 25 & 6.39 & 159 & 0.22 & 0.26 & 0.00 & 20.3 & 18.4 & 1.9 & 2.949 & 0.00 & 2.21$_{1.91}^{2.55}$ & 44.439 & 44.651\smallskip \\
45 & 430 & 9 & 10 53 15.3 & +57 26 30 & 4.91 & 158 & 0.14 & 0.64 & 0.00 & 24.3 & 19.0 & 5.3 & ..... & 21.95$_{21.95}^{22.83}$ & 1.75$_{0.96}^{9.76}$ & ...... & ......\smallskip \\
46 & 0 & 1 & 10 52 36.7 & +57 34 02 & 5.31 & 149 & 0.14 & 0.52 & 0.42 & 24.2 & .... & .... & 0.164 & 21.85$_{21.28}^{22.55}$ & 1.51$_{1.00}^{2.74}$ & 41.020 & 41.831\smallskip \\
47 & 0 & 9 & 10 53 48.1 & +57 28 16 & 8.77 & 148 & 0.18 & 0.42 & 0.00 & 25.5 & 18.5 & 7.0 & ..... & 21.88$_{21.56}^{22.16}$ & 1.83$_{1.17}^{2.82}$ & ...... & ......\smallskip \\
48 & 607 & 9 & 10 52 20.2 & +57 23 06 & 6.46 & 147 & 0.19 & 0.28 & 0.00 & 24.1 & 19.8 & 4.3 & ..... & 20.52$_{19.76}^{21.00}$ & 1.91$_{1.63}^{2.34}$ & ...... & ......\smallskip \\
49 & 0 & 9 & 10 51 11.8 & +57 26 35 & 12.47 & 142 & 0.25 & 0.36 & 0.00 & .... & .... & .... & ..... & 19.90$_{19.76}^{20.85}$ & 1.52$_{1.26}^{1.90}$ & ...... & ......\smallskip \\
50 & 901 & 2 & 10 52 53.0 & +57 28 59 & 1.36 & 138 & 0.05 & 0.33 & 1.25 & 15.3 & 11.9 & 3.4 & 0.204 & 23.64$_{23.37}^{23.82}$ & 2.97$_{1.93}^{4.02}$ & 41.031 & 42.219\smallskip \\
51 & 18 & 1 & 10 52 28.2 & +57 31 05 & 3.03 & 136 & 0.17 & 0.26 & 0.00 & 22.6 & 19.5 & 3.1 & 0.993 & 0.00 & 1.78$_{1.41}^{2.21}$ & 42.952 & 43.198\smallskip \\
52 & 428 & 1 & 10 53 24.6 & +57 28 19 & 5.62 & 132 & 0.24 & 0.41 & 0.00 & 22.4 & 18.8 & 3.6 & 1.598 & 21.20$_{19.76}^{21.73}$ & 1.77$_{1.49}^{2.08}$ & 43.626 & 44.061\smallskip \\
53 & 38 & 1 & 10 53 29.6 & +57 21 04 & 9.96 & 130 & 0.27 & 0.33 & 0.00 & 21.3 & .... & .... & 1.145 & 20.34$_{19.76}^{20.60}$ & 2.18$_{1.93}^{2.58}$ & 43.514 & 43.539\smallskip \\
54 & 804 & 1 & 10 53 12.4 & +57 34 25 & 6.88 & 128 & 0.19 & 0.28 & 0.00 & 22.7 & 19.1 & 3.6 & 1.213 & 0.00 & 2.24$_{1.92}^{2.45}$ & 43.108 & 43.256\smallskip \\
55 & 0 & 9 & 10 54 10.9 & +57 30 57 & 12.00 & 127 & 0.31 & 0.37 & 0.00 & .... & .... & .... & ..... & 19.92$_{19.76}^{20.00}$ & 2.34$_{2.14}^{2.58}$ & ...... & ......\smallskip \\
56 & 0 & 2 & 10 52 51.6 & +57 32 00 & 3.41 & 125 & 0.16 & 0.28 & 0.00 & 21.3 & 16.8 & 4.5 & 0.664 & 21.41$_{20.00}^{21.84}$ & 1.81$_{1.35}^{2.21}$ & 42.538 & 42.996\smallskip \\
57 & 36 & 1 & 10 52 25.7 & +57 23 02 & 6.22 & 125 & 0.22 & 0.27 & 0.00 & 22.5 & 18.6 & 3.9 & 1.524 & 21.34$_{19.76}^{21.65}$ & 2.56$_{2.06}^{2.82}$ & 43.604 & 43.640\smallskip \\
58 & 802 & 5 & 10 52 21.9 & +57 37 34 & 9.22 & 123 & 0.28 & 0.00 & 0.00 & 15.9 & .... & .... & ..... & ...... & ...... & ...... & ......\smallskip \\
59 & 54 & 1 & 10 53 07.7 & +57 15 05 & 14.11 & 122 & 0.21 & 0.28 & 0.00 & .... & .... & .... & 2.416 & 0.00 & 1.66$_{1.49}^{1.74}$ & 44.139 & 44.546\smallskip \\
60 & 0 & 3 & 10 52 47.6 & +57 36 22 & 7.60 & 121 & 0.00 & 0.72 & 1.12 & 17.9 & .... & .... & 0.118 & 23.35$_{22.82}^{23.71}$ & 2.69$_{2.18}^{3.18}$ & 40.228 & 41.566\smallskip \\
61 & 0 & 9 & 10 52 48.5 & +57 41 28 & 12.69 & 118 & 0.25 & 0.26 & 0.00 & .... & .... & .... & ..... & 20.51$_{19.76}^{21.15}$ & 2.08$_{1.66}^{2.73}$ & ...... & ......\smallskip \\
62 & 0 & 9 & 10 53 01.8 & +57 15 00 & 14.03 & 113 & 0.18 & 0.28 & 0.00 & .... & .... & .... & ..... & 20.75$_{19.76}^{21.34}$ & 1.65$_{1.21}^{2.25}$ & ...... & ......\smallskip \\
63 & 0 & 2 & 10 52 52.1 & +57 31 33 & 3.03 & 106 & 0.07 & 0.45 & 0.00 & 19.1 & 16.0 & 3.1 & ..... & 21.97$_{21.40}^{22.39}$ & 1.74$_{0.95}^{2.60}$ & ...... & ......\smallskip \\
64 & 0 & 9 & 10 53 30.5 & +57 25 16 & 7.31 & 104 & 0.15 & 0.00 & 0.00 & 25.9 & 19.7 & 6.2 & ..... & 0.00 & 1.84$_{1.77}^{1.91}$ & ...... & ......\smallskip \\
65 & 0 & 9 & 10 52 11.1 & +57 32 04 & 5.38 & 102 & 0.06 & 0.53 & 0.00 & 25.3 & 19.5 & 5.8 & ..... & 22.04$_{21.65}^{22.42}$ & 1.74$_{1.40}^{2.60}$ & ...... & ......\smallskip \\
66 & 870 & 1 & 10 52 25.4 & +57 22 51 & 6.40 & 101 & 0.18 & 0.00 & 0.00 & 21.3 & 17.8 & 3.5 & 0.807 & 0.00 & 2.25$_{1.89}^{2.69}$ & 42.701 & 42.663\smallskip \\
67 & 34 & 4 & 10 52 58.2 & +57 23 56 & 5.27 & 100 & 0.17 & 0.25 & 0.00 & 26.6 & .... & .... & 0.340 & ...... & ...... & 42.173 & 42.546\smallskip \\
68 & 128 & 4 & 10 53 50.8 & +57 25 13 & 9.81 & 98 & 0.21 & 0.27 & 0.00 & 19.9 & .... & .... & 0.033 & ...... & ...... & 39.754 & 39.902\smallskip \\
69 & 634 & 1 & 10 53 11.6 & +57 23 08 & 6.86 & 97 & 0.13 & 0.23 & 0.00 & 23.2 & .... & .... & 1.544 & 0.00 & 1.43$_{1.16}^{1.73}$ & ...... & 43.715\smallskip \\
70 & 0 & 9 & 10 53 15.5 & +57 24 51 & 5.90 & 95 & 0.04 & 0.53 & 0.00 & 24.5 & 19.4 & 5.1 & ..... & 22.12$_{21.63}^{22.54}$ & 1.77$_{1.02}^{3.09}$ & ...... & ......\smallskip \\
71 & 82 & 1 & 10 53 12.3 & +57 25 06 & 5.41 & 94 & 0.14 & 0.00 & 0.00 & 22.8 & 19.0 & 3.8 & 0.960 & 0.00 & 1.64$_{1.38}^{1.93}$ & 43.022 & 43.471\smallskip \\
72 & 0 & 9 & 10 52 31.8 & +57 24 30 & 4.55 & 94 & 0.00 & 0.49 & 0.34 & 24.1 & 19.6 & 4.5 & ..... & ...... & ...... & ...... & ......\smallskip \\
73 & 19 & 1 & 10 51 37.3 & +57 30 42 & 9.02 & 92 & 0.13 & 0.34 & 0.00 & 22.2 & 19.0 & 3.2 & 0.894 & 20.20$_{19.76}^{21.64}$ & 1.68$_{1.33}^{2.30}$ & 42.521 & 42.843\smallskip \\
74 & 0 & 9 & 10 52 43.4 & +57 35 46 & 6.97 & 91 & 0.14 & 0.00 & 0.00 & .... & .... & .... & ..... & 0.00 & 1.66$_{1.28}^{2.64}$ & ...... & ......\smallskip \\
75 & 0 & 9 & 10 51 46.6 & +57 30 35 & 7.78 & 91 & 0.18 & 0.00 & 0.00 & 25.6 & 19.2 & 6.4 & ..... & 0.00 & 1.96$_{1.38}^{2.38}$ & ...... & ......\smallskip \\
76 & 0 & 9 & 10 51 55.3 & +57 29 34 & 6.45 & 90 & 0.12 & 0.00 & 0.00 & 24.7 & 20.4 & 4.3 & ..... & 0.00 & 1.63$_{1.22}^{2.30}$ & ...... & ......\smallskip \\
77 & 0 & 9 & 10 53 47.1 & +57 33 51 & 9.98 & 90 & 0.21 & 0.00 & 0.00 & .... & .... & .... & ..... & 0.00 & 1.91$_{1.64}^{2.27}$ & ...... & ......\smallskip \\
78 & 861 & 1 & 10 53 58.3 & +57 29 23 & 10.14 & 89 & 0.15 & 0.00 & 0.00 & 22.5 & .... & .... & 1.843 & 0.00 & 1.84$_{1.59}^{2.05}$ & 43.725 & 44.007\smallskip \\
79 & 228 & 4 & 10 53 39.8 & +57 35 18 & 10.01 & 88 & 0.20 & 0.42 & 0.00 & 22.8 & 17.2 & 5.6 & 1.250 & ...... & ...... & 43.843 & 44.022\smallskip \\
80 & 0 & 9 & 10 52 07.2 & +57 34 12 & 7.23 & 87 & 0.12 & 0.27 & 0.00 & 23.1 & .... & .... & ..... & ...... & ...... & ...... & ......\smallskip \\
\vspace{-2 mm} \\
\hline
\vspace{-2 mm} \\
\end{tabular}
}
\normalsize
\end{sidewaystable}
\twocolumn

\onecolumn
\begin{sidewaystable}[h!]
\centering
\bigskip
{\scriptsize
\begin{tabular}{rrcccrrrrrcccccccc}
\multicolumn{18}{l}{\footnotesize {\bf Table 2.}  (continued)} \\
\smallskip \\
\hline
\vspace{-2 mm} \\
XID   & Rosat & Type$^a$ & RA & Dec & Off-axis & counts &  Flux$^b$ & Flux$^b$ & Flux$^b$ & R & K$^\prime$ & R-K$^\prime$ & z & ${\rm log(N_H)}^c$ & $\Gamma^d$ & L$_X$$^e$ & L$_X$$^e$\\
 &  &  &  &  & angle$^f$ & [0.5-7] & [0.5-2] & [2-10] & [5-10] &  &  &  &  &  &  & [0.5-2] & [2-10]\\
\vspace{-2 mm} \\
\hline
\vspace{-2 mm} \\
81 & 0 & 9 & 10 52 54.8 & +57 31 51 & 3.44 & 87 & 0.08 & 0.28 & 0.33 & 22.9 & .... & .... & ..... & 21.71$_{20.00}^{22.24}$ & 1.22$_{0.39}^{1.98}$ & ...... & ......\smallskip \\
82 & 0 & 9 & 10 52 37.0 & +57 16 03 & 12.77 & 86 & 0.15 & 0.00 & 0.00 & .... & .... & .... & ..... & 0.00 & 1.98$_{1.71}^{2.24}$ & ...... & ......\smallskip \\
83 & 0 & 9 & 10 53 30.9 & +57 39 23 & 12.39 & 81 & 0.00 & 0.64 & 0.61 & .... & .... & .... & ..... & 22.04$_{21.95}^{22.70}$ & 1.10$_{0.40}^{1.80}$ & ...... & ......\smallskip \\
84 & 0 & 2 & 10 51 50.1 & +57 25 21 & 7.90 & 80 & 0.07 & 0.34 & 0.00 & 21.8 & 17.4 & 4.4 & 0.676 & 21.97$_{21.18}^{22.37}$ & 1.42$_{0.83}^{2.23}$ & 41.890 & 42.901\smallskip \\
85 & 0 & 9 & 10 53 21.4 & +57 31 48 & 5.98 & 78 & 0.10 & 0.27 & 0.00 & 25.5 & 19.3 & 6.2 & ..... & 21.53$_{20.90}^{21.53}$ & 2.42$_{1.47}^{4.39}$ & ...... & ......\smallskip \\
86 & 0 & 9 & 10 53 46.9 & +57 26 06 & 9.01 & 75 & 0.10 & 0.00 & 0.39 & 23.2 & 19.6 & 3.6 & ..... & 0.00 & 2.00$_{1.52}^{3.03}$ & ...... & ......\smallskip \\
87 & 0 & 9 & 10 52 23.8 & +57 25 32 & 4.16 & 75 & 0.09 & 0.00 & 0.00 & 24.7 & 19.6 & 5.1 & ..... & 0.00 & 1.63$_{1.26}^{2.37}$ & ...... & ......\smallskip \\
88 & 41 & 4 & 10 53 19.0 & +57 20 48 & 9.36 & 73 & 0.15 & 0.00 & 0.00 & 17.5 & .... & .... & 0.340 & ...... & ...... & 42.336 & 42.279\smallskip \\
89 & 0 & 9 & 10 52 42.3 & +57 29 11 & 0.41 & 73 & 0.08 & 0.20 & 0.00 & 24.0 & 19.8 & 4.2 & ..... & 19.60$_{19.76}^{21.57}$ & 1.06$_{0.68}^{1.65}$ & ...... & ......\smallskip \\
90 & 33 & 1 & 10 52 00.3 & +57 24 20 & 7.27 & 73 & 0.07 & 0.20 & 0.32 & 22.2 & 19.4 & 2.8 & 0.974 & 0.00 & 0.67$_{0.20}^{1.37}$ & 42.309 & 43.115\smallskip \\
91 & 0 & 9 & 10 51 20.6 & +57 26 59 & 11.22 & 73 & 0.12 & 0.00 & 0.00 & .... & .... & .... & ..... & ...... & ...... & ...... & ......\smallskip \\
92 & 0 & 3 & 10 52 58.6 & +57 33 35 & 5.24 & 72 & 0.10 & 0.00 & 0.00 & 24.3 & 19.0 & 5.3 & 0.417 & 20.97$_{20.90}^{22.52}$ & 1.61$_{1.10}^{2.53}$ & 41.853 & 42.282\smallskip \\
93 & 21 & 9 & 10 51 55.1 & +57 30 43 & 6.70 & 72 & 0.09 & 0.00 & 0.00 & 24.2 & 20.1 & 4.1 & ..... & 0.00 & 1.89$_{1.30}^{2.90}$ & ...... & ......\smallskip \\
94 & 434 & 2 & 10 52 58.4 & +57 22 51 & 6.30 & 71 & 0.13 & 0.00 & 0.00 & 22.2 & 18.0 & 4.2 & 0.762 & 21.36$_{20.78}^{21.66}$ & 2.12$_{1.69}^{2.56}$ & 42.862 & 43.123\smallskip \\
95 & 0 & 9 & 10 52 31.5 & +57 25 03 & 4.05 & 71 & 0.00 & 0.23 & 0.38 & 24.8 & 19.8 & 5.0 & ..... & 22.28$_{19.76}^{22.85}$ & 1.13$_{-0.06}^{2.31}$ & ...... & ......\smallskip \\
96 & 39 & 1 & 10 52 09.7 & +57 21 04 & 8.94 & 71 & 0.14 & 0.00 & 0.00 & 21.7 & 20.6 & 1.1 & 3.281 & 22.69$_{22.30}^{23.00}$ & 2.55$_{2.00}^{3.20}$ & ...... & 44.472\smallskip \\
97 & 108 & 2 & 10 52 27.8 & +57 33 30 & 5.14 & 71 & 0.09 & 0.00 & 0.00 & 22.1 & 18.4 & 3.7 & 0.696 & 21.08$_{19.76}^{22.01}$ & 1.36$_{0.78}^{2.33}$ & 42.097 & 42.719\smallskip \\
98 & 0 & 9 & 10 52 41.5 & +57 30 39 & 1.88 & 70 & 0.07 & 0.00 & 0.00 & 25.5 & 20.3 & 5.2 & ..... & 0.00 & 1.46$_{1.21}^{1.51}$ & ...... & ......\smallskip \\
\vspace{-2 mm} \\
\hline
\vspace{-2 mm} \\
\multicolumn{18}{l}
{$^a$ Optical classification: 1=Type1 AGN, 2=Type2 AGN, 3= galaxy, 4=group/cluster, 5=star, 9=unidentified source} \\
\multicolumn{18}{l}
{$^b$ In units of $10^{-14}$ erg s$^{-1}$.} \\
\multicolumn{18}{l}
{$^c$ Intrinsic absorption in excess to the Galactic ones ($\sim 5.7\times 10^{19}$ $cm^{-2}$) and 90$\%$ confidence range.} \\
\multicolumn{18}{l}
{$^d$ Spectral Index and $90\%$ confidence range.} \\
\multicolumn{18}{l}
{$^e$ Log of observed luminosity in the rest-frame band, in units of erg s$^{-1}$.} \\
\multicolumn{18}{l}
{$^f$ In arcmin.} \\
\multicolumn{18}{l}
{$^g$ Photometric redshifts (see Lehmann et al. \cite{ingo01a}).} \\
\end{tabular}
}
\normalsize
\end{sidewaystable}
\twocolumn

\end{document}